\documentclass[twocolumn,prb,amsmath,showpacs,amssymb,citeautoscript,floatfix]{revtex4}

\usepackage{graphicx,color}

\newcommand{\tcb}{\textcolor{black}}

\usepackage{dcolumn}
\usepackage{bm,ulem}
\usepackage{amsmath,amssymb}
\begin{document}

\title{Physical properties of the tetragonal CuMnAs: a first-principles study}

\author{F. M\'aca}\email{maca@fzu.cz} \affiliation{Institute of Physics ASCR, Na Slovance 2,
CZ-182 21 Praha 8, Czech Republic}

\author{J. Kudrnovsk\'y} \affiliation{Institute of Physics ASCR, Na Slovance 2, CZ-182 21 Praha 8,
Czech Republic}

\author{V. Drchal} \affiliation{Institute of Physics ASCR, Na Slovance 2, CZ-182 21 Praha 8, Czech
Republic}

\author{K. Carva} \affiliation{Charles University, Faculty of Mathematics and Physics, Department
of Condensed Matter Physics, Ke Karlovu 5, CZ-121 16 Praha 2, Czech Republic}

\author{P. Bal\'a\v{z}} \affiliation{Charles University, Faculty of Mathematics and Physics,
Department of Condensed Matter Physics, Ke Karlovu 5, CZ-121 16 Praha 2, Czech Republic}

\author{I. Turek} \affiliation{Charles University, Faculty of Mathematics and Physics, Department
of Condensed Matter Physics, Ke Karlovu 5, CZ-121 16 Praha 2, Czech Republic} \date{\today}

\begin{abstract}

Electronic, magnetic, and transport properties of the antiferromagnetic (AFM) CuMnAs alloy with
tetragonal structure, promising for the AFM spintronics, are studied from first principles using
the Vienna ab-initio simulation package. We investigate the site-occupation of
sublattices and the lattice parameters of three competing phases. We analyze the factors that
determine which of the three conceivable structures will prevail. We then estimate formation
energies of possible defects for the experimentally prepared lattice structure. Mn$_{\rm Cu}$- and
Cu$_{\rm Mn}$-antisites as well as Mn$\leftrightarrow$Cu swaps and vacancies on Mn or Cu sublattices
were identified as possible candidates for defects in CuMnAs. We find that the interactions of the
growing thin film with the substrate and with vacuum as well as the electron correlations are
important for the phase stability while the effect of defects is weak. In the next step, using the
tight-binding linear muffin-tin orbital method for the experimental structure, we estimate
transport properties for systems containing defects with low formation energies. Finally, we determine the exchange
interactions and estimate the N\'eel temperature of the AFM-CuMnAs alloy using the Monte Carlo
approach. A good agreement of the calculated resistivity and N\'eel temperature with experimental
data makes possible to draw conclusions concerning the competing phases.

\end{abstract}


\maketitle

\section{Introduction}

The tetragonal antiferromagnetic (AFM) CuMnAs phase prepared by the molecular-beam epitaxy (MBE)
on the GaAs(001) and GaP(001) substrates has attracted recently considerable experimental and
theoretical interest in connection with the so-called AFM
spintronics.\cite{afm-spin,cumnas-1,cumnas-2,cumnas-3} The combined experimental and theoretical
study \cite{cumnas-1} (see also Ref. ~\onlinecite{cumnas-str}) has lead to a proposal of basic
structural parameters that were used in first-principles calculations assuming an ideal structure
without defects. \cite{cumnas-1} On the other hand, the experiment for this phase provides the
basic physical parameters: the residual resistivity around 90~$\mu\Omega$cm for $T=5$~K
(Ref. ~\onlinecite{cumnas-1}), the N\'eel temperature around 480~K (Ref. ~\onlinecite{cumnas-tn}),
and the local Mn moments around 3.6~$\mu_{\rm B}$ at room temperature (Ref.
 ~\onlinecite{cumnas-1}); its value at lower temperature will be higher (see also
Sec.~\ref{EI-cumnas}). The transport studies (the residual resistivity) thus indicate the presence
of defects whose origin and concentrations are known only very approximately (sample grown on the
GaAs substrate). \cite{cumnas-str} Identification of possible defects and their formation energies
thus represent a challenge for the theory. The same concerns also an estimate of the residual
resistivity and the N\'eel temperature. The N\'eel temperature is closely related to corresponding
exchange interactions and, in turn, also to the values of local Mn magnetic moments.

Another interesting issue concerns the sample preparation. The CuMnAs in the bulk phase
crystallizes in the orthorhombic phase \cite{cumnas-or} while the studied tetragonal phase does
not \tcb{exist as a bulk phase in the Nature} and can be only prepared as a film by the MBE on a suitable substrate.

The aim of the present study is thus two-fold.
First, we will determine theoretically the structure of the tetragonal
phase by optimizing the lattice parameters ($a$=$b$ and $c$ in
the present case) and positions of Cu, Mn, and As atoms inside the
unit cell.
We also investigate the effect of the substrate, defects, and of electron
correlations \cite{afm-mnte} on the phase stability. Moreover, we
estimate the formation energies of possible defects.
Second, we will calculate relevant physical quantities such as the
local Mn-moments, exchange interactions, and the N\'eel temperature as
well as the residual resistivity due to specific defects. These
quantities will be determined in the framework of the unified
first-principle electronic structure model and compared with the
experiment.

\section{Formalism} \label{Form}

\begin{figure}
\center \includegraphics[width=5cm]{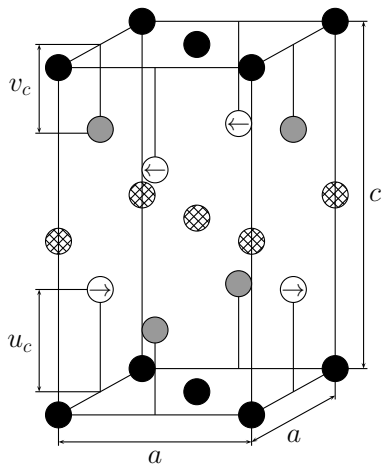}
\caption { The lattice structure of the phase I of tetragonal AFM CuMnAs
consisting of two nonmagnetic Cu-/As-sublattices (black/grey color), and
two Mn-sublattices with the antiparallel spin orientations (indicated by arrows).
The possible interstitial positions (Wyckoff positions 2b) are shown by
hatched circles.
For the phase II the positions of Cu- and Mn-atoms are interchanged.} \label{f1}
\end{figure}

The AFM-CuMnAs prepared by the MBE has a tetragonal structure \cite{cumnas-1,cu2sb-str} with the
space group P4/nmm (No.~129). \cite{IntT-A}
The experimental lattice parameters are $a$=$b$=3.82~\AA~and $c$=6.318~\AA.
The atomic basis contains two formula units (6 atoms), Cu atoms are in the basal
plane of the tetragonal lattice, (Wyckoff position $2a$), there are two
parallel layers of As atoms (Wyckoff position $2c$) and two layers of Mn atoms
(Wyckoff position $2c$) with oppositely oriented moments (see Fig.~\ref{f1}).
The interstial sites are located in the Wyckoff positions $2b$.
The relative positions of atoms (in units of $c$) are $z_{\rm Cu}$=0.0,
$z_{\rm Mn}$=$u_{c}$=0.330, and $z_{\rm As}$=$v_{c}$=0.266.
We call this structure phase I.
In the other possible structure, which we call phase II, Mn atoms are in
position $2a$ and Cu and As atoms occupy positions $2c$.
Finally, there could be a structure (denoted as phase III) with As atoms
in the basal plane ($2a$) and Cu and Mn atoms in positions $2c$.

Theoretical lattice parameters $a$ and $c$, as well as atomic positions in the unit cell are
determined by the VASP calculations (Vienna ab-initio simulation package using the projector
augmented wave scheme \cite{paw}) with different exchange correlation potentials, namely, the LDA
(Vosko-Wilk-Nusair, VWN) \cite{VWN}, the GGA (Perdew-Burke-Ernzerhof, PBE) \cite{GGA}, and the
GGA+U with a simple empirical on-site Coulomb interaction $U$ related to Mn-$d$ orbitals.
\cite{ldau} The supercell VASP calculations (48 atoms) is used to
determine formation energies of possible simple defects assuming the
experimental lattice structure.
\tcb{For VASP calculations we have used plane-waves up to 350~eV
and the Brillouin zone sampling with 270 special k-points in the irreducible
three-dimensional wedge and corresponding number of k-points in
the supercell.}

The transport coefficients, exchange interactions, and the N\'eel temperature are determined using
the Green function formulation of the tight-binding linear muffin-tin orbital (TB-LMTO) method in
which the effect of disorder (defects) is described by the coherent potential approximation (CPA).
\cite{book} The TB-LMTO method employs the atomic sphere approximation (ASA) and it is thus less
accurate than the VASP technique. We have therefore compared relevant electronic properties (local
moments and densities of states (DOS)) with the VASP results for the ideal, defect-free AFM-CuMnAs
assuming the experimental structure. This is an important check for more complex, non-cubic
structures (see, e.g., Ref. ~\onlinecite{bi2te3}). Calculations are done using the VWN
exchange-correlation potential, but we also check the robustness of result with respect to the
electron correlations (on-site Coulomb interaction model). We have neglected the spin-orbit
effects in both approaches.

The transport studies employ the Kubo-Greenwood linear response theory in which the
disorder-induced vertex-corrections are included in the CPA. \cite{vertex} Their inclusion is
simplified by the present formulation of the velocity as the intersite hopping \cite{rho-our}
which leads to non-random velocity matrices.

The effective exchange interactions between Mn atoms for a given shell $s$,
$J_{s}$, are determined by the Liechtenstein mapping procedure \cite{lie}
generalized to random alloys.  \cite{iec-our}
\tcb{ As a result, we obtain the effective Heisenberg Hamiltonian
$H = - \sum_{ij} J_{ij} \mathbf{e}_i \cdot \mathbf{e}_j$  which
will be used for the estimate of the N\'eel temperature.
The indices $i$ and $j$ run over all sites occupied by Mn atoms,
$J_{ij}$ denote the pair exchange interactions and the unit vectors
$\mathbf{e}_i$ define the local moment directions.
We note that the positive/negative values
of $J_{s}$ correspond to the ferromagnetic (FM)/ AFM interactions
and that the values of magnetic moments are included in their
definitions.}\cite{lie,iec-our}
The exchange interactions depend on the reference magnetic state from
which they are
extracted. In particular, in the present case we employ two reference states, namely, the AFM
state and the disordered local moment (DLM) state. \cite{dlm} The DLM approach describes the
paramagnetic state above the critical temperature with fluctuating Mn-moments and it is thus
better suited for the estimate of the critical (N\'eel) temperature for the transition between the
AFM and paramagnetic states as compared to the AFM reference state corresponding to zero
temperature. The DLM state is treated as a random equiconcentration binary alloy of moments
pointing randomly in opposite directions and can be thus naturally treated using the CPA.
\cite{dlm} In both cases, however, local moments on Cu- and As-atoms are strictly zero. It should
be noted that due to the two Mn-sublattices we will have two sets of exchange parameters, intra-
and inter-sublattice ones. We remark that the DLM state will be also used in transport studies.

To study the thermodynamic properties of CuMnAs we employed classical
Monte Carlo (MC) simulations based on the Metropolis algorithm \cite{binder}
applied to the constructed Heisenberg Hamiltonian.
 For simulations we used a three-dimensiona supercell composed of
16$\times$16$\times$16 elementary CuMnAs cells with periodic boundary
conditions.
The simulations were carried out assuming zero applied magnetic field
and disregarding magnetocrystalline anisotropy.
The local Mn magnetic moments, as large as $3.80\, \mu_{\rm B}$, were
assumed to be independent of temperature.
We started the simulation from an initial temperature $900\,{\rm K}$
which decreased by a step $\Delta{T} = 10\, {\rm K}$.
At each temperature $2\times 10^5$ MC steps were performed.
To accumulate the statistics we simultaneously simulated 5 independent
identical systems.

\section{Results and discussion}

\subsection{Ground state of the ideal tetragonal AFM-CuMnAs} \label{GS-cumnas}

Assuming the ideal tetragonal AFM-CuMnAs, two natural questions arise: (i) the occupation of
atomic positions inside the elementary cell by Cu, Mn, and As atoms, i.e., which phase (I, II or
III) corresponds to the ground state; and (ii) optimal lattice parameters $a$=$b$, and $c$ and the
coordinates of atoms inside the unit cell.
This structure allows both the FM alignment (like, e.g., in CrMnAs \cite{cu2sb-str})
and the AFM-alignment (like in CuMnAs), but also more complex magnetic structures could exist,
if one considers larger unit cells, but here we limit ourselves
to the case of six atoms per unit cell observed in the experiment.\cite{cumnas-1}

Results of extensive calculations in the framework of the VASP and GGA-PBE are summarized in
Table~\ref{t1} for the optimized phases I and II together with distances between atoms. We have
also included results for the experimental geometry. \cite{cumnas-1}
\begin{table}[h]
\caption{ Total energies per elementary cell of the phase II with respect to the phase I (energy
zero) are shown assuming the (frozen) experimental geometry (label 0) as well as the optimized
one. Also shown are lattice parameters $a$, $c$, the relative $z$-coordinates of Cu, Mn, As atoms
inside the unit cell, its volume $V$, and the local Mn-moments $m^{\rm Mn}$. In the bottom part of
the Table the nearest-neighbor distances between various atom pairs in all structures are given.
\\}
\renewcommand{\arraystretch}{1.2 }
\tcb{\begin{tabular}{ccccc}\hline\hline
 ~& ~phase I$_{\rm 0}$~ & ~phase I~ & ~phase II$_{\rm 0}$~ & ~phase II~ \\ \hline
 $a$ [\AA]&3.82&3.69&3.82&3.85\\
 $c$ [\AA]&6.32&6.40&6.32&5.94\\
 $V$ [\AA$^3$]&92.20&87.14&92.20&88.05\\
 $z_{\rm Cu}$&0.00&0.00&0.670&0.682\\
 $z_{\rm Mn}$&0.670&0.651&0.00&0.00\\
 $z_{\rm As}$&0.266&0.273&0.266&0.270\\
 $m^{\rm Mn}$ [$\mu_B$]&3.70&3.41&2.96&2.78\\
 $\Delta E_{\rm tot}$ [eV] &0.0&0.0&$-$0.078&$-$0.102\\\hline
 $d_{\rm Cu-Cu}$ [\AA]&2.70&2.61&3.45&3.48\\
 $d_{\rm Cu-Mn}$ [\AA]&2.82&2.90&2.82&2.70\\
 $d_{\rm Cu-As}$ ~[\AA]&2.55&2.54&2.55&2.74\\
 $d_{\rm Mn-As}$ [\AA]&2.55&2.66&2.55&2.51\\
 $d_{\rm Mn-Mn}$ [\AA]&3.45&3.24&2.70&2.72\\
 $d_{\rm As-As}$ ~[\AA]&3.82&3.69&3.82&3.85\\\hline\hline
 \end{tabular}\renewcommand{\arraystretch}{1.0}}
\label{t1}
\end{table}

The following conclusions are made:
(i) The ground state is the phase II, but with the energy preference with
respect to the phase I being only about 0.1~eV per unit cell.
The lattice parameter $c$ is about 6\% smaller as compared to the grown
sample while the lattice parameter $a$ is similar;
(ii) In the phase I the result is just opposite: the lattice parameter $c$
is similar to that in the grown sample, but the lattice parameter $a$ is
smaller by 3.5\%;
(iii) Theoretical volumes for phases I and II were smaller as compared to
the experimental one thus indicating a possible role played by the substrate;
(iv) The energy preference of the phase II as compared to the phase I (by about
0.08~eV per unit cell) is obtained also for the experimental structure;
(v) The values of local Mn moments are strongly underestimated in both
phases II and III as compared to experiment; and
(vi) The total energy of the phase III (with As atoms in the basal plane)
was estimated for the experimental lattice parameters, but with optimized
atom positions.
\tcb{It was higher than that of the phase I by 2.97~eV.}
Calculated interatomic distances among atoms (Table~\ref{t1}) also indicate a
possible experimental test -- using the Extended X-ray Absorption Fine Structure (EXAFS)
experiment which could distinguish between possible phases, namely, by checking the
nearest-neighbor Mn-Mn distances which differ significantly and do not interfere
with distances between other atom pairs.
Additional arguments in favor of the phase I will be given below based on the
transport studies and an estimate of the N\'eel temperature.

\tcb{For the phase I at the experimental geometry we have also
estimated total energies of the FM and non-magnetic CuMnAs phases (+0.29 eV
and +2.82 eV), respectively.
Corresponding total energies are higher as
compared to the AFM total energies so that they can be excluded
as possible ground state candidates.
On the other hand, the energy difference between the AFM and FM
states is smaller for the phase~II as
compared to phase~I (+0.069~eV and +0.29~eV, respectively).}
Such result is compatible with exchange interactions of both
phases~I and II, namely with dominating AFM interactions
for the former and competing FM and AFM interactions for the
latter (see Figs.~\ref{f4}b and \ref{f4}c below).

While the semilocal GGA exchange-correlation potential is generally considered to be an optimal
choice for the structure optimization, the GGA+U approach is sometimes used to fine 
the theoretical description in some systems, tuning of band gaps, magnetic moments, the critical
temperatures, etc. (see, e.g. a recent study of the AFM-MnTe). \cite{afm-mnte} We present here a
similar study of the effect of electron correlations on the lattice structure and magnetic moments
in the AFM-CuMnAs assuming that the Hubbard parameter $U$ is limited to $d$-orbitals of Mn atoms,
which is an acceptable model for narrow Mn-bands (Table~\ref{t2}).
\begin{table}[h]
\caption{Total energy differences (per elementary cell) between the phases I$_0$ and II$_0$,
 $\Delta E_{tot} = E_{\rm II_0} - E_{\rm I_0}$,
 as a function of the on-site Hubbard parameter $U$.
 Also shown are corresponding local Mn-moments.\\}
\renewcommand{\arraystretch}{1.2}
 \begin{tabular}{ccccc} \hline\hline
  ~~~$U$ [eV]~~~&0&0.41&0.83&1.25\\\hline
  $m^{\rm Mn}_{I_0}$ [$\mu_B$]&~~~~~3.70~~~~&~~~~3.80~~~~&~~~~3.90~~~~&~~~~3.99~~~~\\
  $m^{\rm Mn}_{II_0}$ [$\mu_B$]&~~~~~2.96~~~~&~~~~3.15~~~~&~~~~3.31~~~~&~~~~3.46~~~~\\
  ~$\Delta E_{\rm tot} $[eV]~&$-$0.078&+0.102&+0.278&+0.443\\ \hline\hline
  \end{tabular}\renewcommand{\arraystretch}{1.}
\label{t2}
\end{table}

Electron correlations stabilize the phase I$_0$ as compared to the phase II$_0$.
Already for $U$=0.4~eV the phase I$_0$ has a lower total energy than the phase II$_0$.
Assuming $U$ around 1~eV, the local
Mn-moment (3.9 $-$ 4.0~$\mu_{\rm B}$) agrees also reasonably well considering the fact that it was
measured at room temperature.
We have as well tested the effect of lattice relaxations and found only slight quantitative modifications
 not changing the qualitative picture.

There are two other effects which could influence the calculated phase stability, namely, that
samples are grown on the particular substrate and the presence of impurities in the sample. The
real samples are grown on the As-/P-terminated GaAs(001)/GaP(001) faces. \cite{growth} We have
tried to elucidate a possible role of the substrate using a simple model which simulates this
case, namely, the system consisting of five layers of GaP simulating the substrate with an
extra layer of P atoms (P-rich conditions) which interface with four multilayers of CuMnAs,
either in the phase I (the bottom layer is Cu one) or in the phase II  structure (the bottom layer
is Mn one). Such system separated by a vacuum layer is periodically repeated and studied by the
supercell method. We used the VASP-GGA, fixed the substrate layers, but allowed relaxation of
extra layer of P atoms, and atoms inside CuMnAs. We also varied spin orientations, the AFM
orientation was either between Mn-layers or inside each Mn-layer. In all cases, the AFM
orientation between Mn-layers was preferred. Finally, we have also tested models with frozen
experimental sample geometry. In all cases we have obtained the preference of the phase I, the
energy difference in its favor was quite substantial and varied between 0.8 and 1 eV per
elementary cell with 42 atoms.

The present model fulfills the basic requirement for comparison of total energies, namely, the
same number of atoms in supercells. The model correctly includes sample/vacuum and
sample/substrate interfaces as it is in a real system. On the other hand, it has certain
limitations as we do not consider possible switching between the phases I and II during the
growth. One should keep in mind that theoretical calculations assume zero temperature and give the
global minimum of energy while a real sample exists in a non-equilibrium state due to sample
preparation and it can be in the local energy minimum.

We refer the reader to the end of the next Section as concerns a possible effect of impurities on
the phase stability.

\subsection{Formation energies of defects in AFM-CuMnAs} \label{FE-cumnas}

Structural study \cite{cumnas-str} and measurements of residual resistivity (90~$\mu\Omega$cm)
indicate that the samples contain defects. An estimate of the formation energies (FE) of defects
is a tool that can identify possible candidates. A complete study of all possible defects, similar
to that done for a cubic FM-NiMnSb \cite{nimnsb-fe} is beyond the scope of the present paper.
Rather, we choose a few possible candidates as in the study for related CuMnSb alloy
\cite{cumnsb-fe}. We have estimated FE for chosen substitutional defects including vacancies, as
well as for Mn-interstitial and listed them in Table~\ref{t3}.
\begin{table}[h]
\caption{
The formation energies FE for various substitutional defects in the
tetragonal AFM-CuMnAs.
Also studied was the Mn-interstitial (Mn$_{\rm int}$, see Fig.~\ref{f1}).
The symbol X$_{\rm Y}$ denotes the X-defect on the Y-sublattice.
Defects are sorted according to their formation energies, the values for
unrelaxed atom positions are given in brackets  . \\}
\renewcommand{\arraystretch}{1.2}
\begin{tabular}{cccc}\hline\hline
~~~Defect~~~ & ~~~FE [eV]~~~ & ~~~~~Defect~~~~~ & ~~~FE [eV]~~~ \\\hline
Vac$_{\rm Mn}$&$-$0.13 ($-$0.23)&Mn$_{\rm int}$ &+1.62 (+2.15)\\
Vac$_{\rm Cu}$&$-$0.13 ($-$0.10)&As$_{\rm Cu}$&+1.73 (+2.66)\\
Mn$_{\rm Cu}$&$-$0.04 ($-$0.06)&As$_{\rm Mn}$&+1.77 (+1.90)\\
Cu$_{\rm Mn}$&+0.33 (+0.27)&Mn$_{\rm As}$&+2.00 (+1.95)\\
Cu$_{\rm As}$&+1.15 (+1.06)&Vac$_{\rm As}$&+2.18 (+2.22)\\ \hline\hline
\end{tabular}\renewcommand{\arraystretch}{1.}
\label{t3}
\end{table}

The supercell VASP method
and GGA-PBE was applied to the reference 48-atoms supercell Cu$_{16}$Mn$_{16}$As$_{16}$ and to
corresponding supercells containing specific defects. For example, Cu$_{15}$Mn$_{17}$As$_{16}$
supercell simulates the Mn$_{\rm Cu}$ defect concentration of 6.25\%. We have used the
experimental lattice parameters. The
accurate determination of FE is a challenging task (see, e.g., a recent review \cite{theory-fe}).
Here we employ the simplest possible approach in which the FE is defined as FE=$E_{\rm tot}$[def]
$- E_{\rm tot}$[id] $- \sum_{i} n_{i} E_i$, where $E_{\rm tot}$[def] and $E_{\rm tot}$[id] are
total energies of the supercells with (def) and without (id) defects, $n_{i}$ indicates the number
of atoms of type $i$ ($i$=Cu, Mn, As, vacancy) that have been added to ($n_{i} > 0$) or removed
from ($n_{i} < 0$) the supercell when the defect is formed, and $E_{i}$ are total energies of
atoms in their most probable bulk phase. \cite{theory-fe}
\tcb{ Strictly speaking, one should employ instead of $E_i$  corresponding
chemical potentials of these species which may depend on the temperature,
defect concentration, the presence of other defects, etc.
The above choice represents a rough, but acceptable approximation. }
It was used, e.g., in Refs. ~\onlinecite{nimnsb-fe, cumnsb-fe} for cubic
semi-Heusler NiMnSb and CuMnSb alloys. We have chosen for $E_{i}$ the total energies of fcc-Cu,
AFM-Mn (L1$_{0}$-lattice), and rhombohedral As. The choice for Mn is the same as in
Ref. ~\onlinecite{nimnsb-fe} although in the OQMD (Open Quantum Materials Database \cite{oqmd-fe})
a more complex structure is used \cite{rem-mn}. It should be noted that actual values for the FE
may depend on the choice of these energies and on the determination of $E_{\rm tot}$[def]. While
the lattice parameters ($a$, $c$) were kept fixed in all cases we have optimized atomic positions
inside the supercell. We have also tested the model with frozen atomic positions like in the ideal
structure, but there were only small quantitative differences. Results are summarized in
Table~\ref{t3} with the following conclusions: (i) Mn$_{\rm Cu}$ and Cu$_{\rm Mn}$ are similarly
as in the cubic CuMnSb the most probable candidates for possible defects. In addition, also
vacancies on Mn- and Cu-sublattices have small FE. We note that a small FE for Mn-vacancy
\cite{nimnsb-fe} was also found for NiMnSb alloy; (ii) Mn-interstitials have, contrary to CuMnSb
or NiMnSb, a much larger FE due to tetragonal vs cubic structure with natural vacancy sites in the
latter; (iii) Also Mn$_{\rm As}$ or As$_{\rm Mn}$ and related defects have large FE similarly like
Mn$_{\rm Sb}$  and Sb$_{\rm Mn}$ in NiMnSb or CuMnSb \cite{nimnsb-fe,cumnsb-fe}; and (iv) Although
the FE of Mn$\leftrightarrow$Cu swaps was not explicitly studied, one can roughly estimate it as
the sum of FE's of Mn$_{\rm Cu}$ and Cu$_{\rm Mn}$ assuming that they are not correlated.
\cite{cumnsb-fe} Consequently, the Mn$\leftrightarrow$Cu swaps are also possible candidates.
\tcb{To resume, we regard the defects with FE in tenths of eV as probable, while those with
FE above \mbox{1~eV} we do not further consider.}

Essentially zero (small negative) FE for Mn-Cu swap correlates with the fact that corresponding
total energies of phases I and II have very similar energies (see Table~\ref{t1}), the phase II
has the total energy even slightly lower. Besides the FE's the formation of defects depends on
delicate details of the impurity kinetics which is not considered here. Anyway, defects with low
FE's are more probable candidates than those with larger FE's even at the non-equilibrium
conditions. We will therefore investigate below the influence of these more probable defects on
transport properties. It should be noted that defects play less important role on the value of the
N\'eel temperature as compared to the resistivity so that we limit ourselves to defect-free
samples when calculating N\'eel temperature.

We have further studied possible effect of impurities on the phase stability. We have taken two
defect types with low and high formation energies (see below), namely Mn$_{\rm Cu}$ and Mn$_{\rm
As}$ and investigated also the effect of the defect concentrations (supercells simulating the
defect concentrations 6.25\% and 12.5\%, respectively). Neither the defect type nor the higher
defect concentration were able to change the energy preference of the phase II.

\subsection{Transport properties of AFM-CuMnAs} \label{RHO-cumnas}

We will estimate residual resistivities due to possible defect types found in the previous Section, namely,
Mn$_{\rm Cu}$, Cu$_{\rm Mn}$, Mn-vacancy, and Mn$\leftrightarrow$Cu swap assuming the defect
concentration of 5\% in each case.

To this end we employ the linear-response theory as formulated in the TB-LMTO-CPA method
\cite{rho-our} and including disorder-induced vertex-corrections \cite{vertex} and neglect
possible relativistic effects (spin-orbit interaction) for simplicity.

\begin{figure}
\center \includegraphics[width=7.cm]{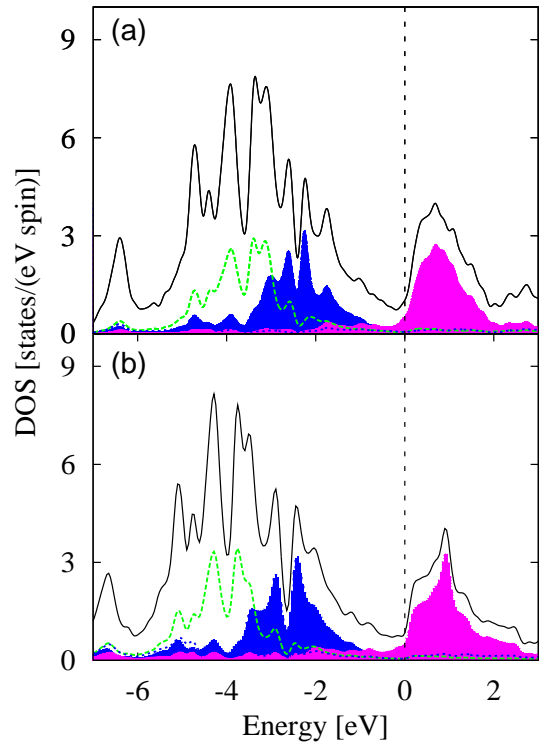}
\caption {(Color online) Comparison of total and
local densities of states (DOS) for the phase I of AFM-CuMnAs alloy evaluated using (a) VASP
method and (b) the TB-LMTO method in the LDA framework. The spin-resolved local Mn-DOS's are shown
(majority spin - blue, minority spin - red).  The local Cu- (dashed line) and As-DOS (dotted line) are spin-independent.}
\label{f2}
\end{figure}
We will first
demonstrate that the present tetragonal AFM-CuMnAs alloy can be described properly by the TB-LMTO
method similarly as we did recently \cite{bi2te3} for topological insulator Bi$_{2}$Te$_{3}$.
Careful tests have lead to the conclusion that we can use a model without empty spheres and
assuming the same atomic Wigner-Seitz radii. We present in Figs.~\ref{f2}a and \ref{f2}b the
densities of states (DOS's) for an ideal tetragonal AFM-CuMnAs as calculated by the VASP and
the TB-LMTO methods, respectively, using the VWN exchange-correlation potential in both cases. A very
good agreement between both DOS's is obtained. A similarly good agreement was obtained also for
the phase II and for models with empirical Hubbard $U$ (not shown).
\begin{figure}
\center \includegraphics[width=7cm]{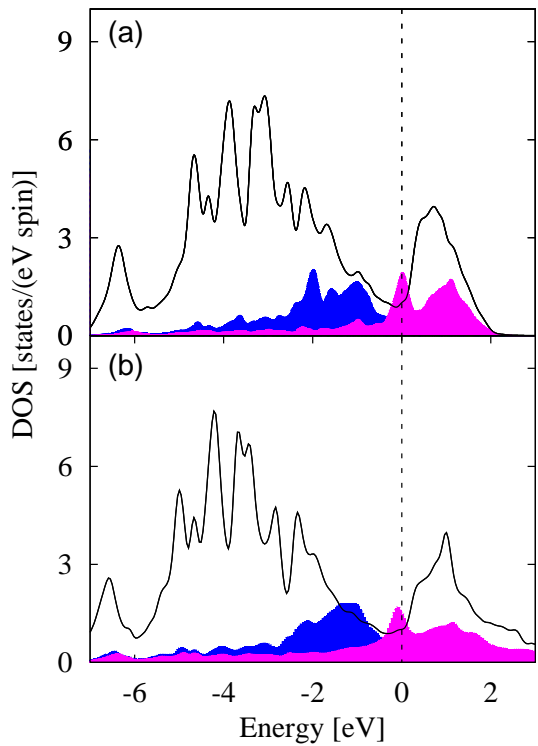}
\caption {(Color online) The total and local densities of
states (DOS) for the reference AFM-CuMnAs alloy with 5\% of extra Mn-atoms on Cu-sublattices (a)
VASP result, (b) LMTO.  We show only the local Mn-DOS's on Cu-sublattice (majority spin - blue, minority spin - red) to show
the pronounced virtual bound state at the Fermi energy in minority states. } \label{f3}
\end{figure}

Concerning the transport properties, there is an important difference between
Mn$_{\rm Cu}$ defects or Mn$\leftrightarrow$Cu swaps on one side and
Cu$_{\rm Mn}$ defects or Mn vacancies on the other hand.
First, the frustration of the Mn$_{\rm Cu}$ moments is obvious (see Fig.~\ref{f1}).
We have therefore considered two limiting models:
(i) A collinear moment alignments, both the parallel or antiparallel (P/AP)
to the nearest native Mn-sublattices, which have the same total energy, and
(ii) the DLM state applied to Mn$_{\rm Cu}$ moments characterizing an ideal
frustrated state.
We have found (5\% defects) that the total energy for the DLM-Mn$_{\rm Cu}$ is
smaller by negligible 0.17~meV per formula unit.
Second, a virtual bound state (VBS) is present at the Fermi energy for
Mn$_{\rm Cu}$ and Mn$\leftrightarrow$Cu swap defects and it is missing
for Cu$_{\rm Mn}$ defects or Mn-vacancies.

The VBS in the DOS for Mn$_{\rm Cu}$ defect is shown in Fig.~\ref{f3}a
and \ref{f3}b for the VASP and TB-LMTO methods, respectively.
We have assumed the DLM-Mn$_{\rm Cu}$ model for the TB-LMTO while for the VASP
we used again 48-atom supercell with a single Mn-atom on the Cu-site
(the concentration 6.25\%).
It should be noted that a similar VBS was found in the TB-LMTO approach also
for the collinear P/AP model and/or for the Mn$\leftrightarrow$Cu swap
model (not shown).
The presence of impurity states at the Fermi energy will lead to a
stronger scattering due to the VBS (Mn$_{\rm Cu}$ or Mn$\leftrightarrow$Cu defects) and
thus larger resistivity (for a comparable defect concentrations) as compared
to the Cu$_{\rm Mn}$ defect or Mn-vacancy.

We have verified that the VBS exists at the Fermi energy also for the LDA+U model. The final
remark is related to the transport geometry, namely, to the fact that the current can flow either
in the ($x$,$y$)-plane (in-plane current), used in the experiment, or normal to it, i.e., in the
$z$-direction (out-of-plane current). We have summarized some typical results for different
defects in Table~\ref{t4}.
\begin{table}[h]
\caption{ The calculated resistivities (in $\mu\Omega$cm) for tetragonal
AFM-CuMnAs with 5\% of different defect types (Models A to G). \tcb{The
resistivity of the paramagnetic state (SDR) is also shown (Model G). }
The experimental values for the sample with unspecified amount of defects
\cite{cumnas-1} are about 90/160~$\mu\Omega$cm as measured at temperatures
$T$=5~K/300~K, respectively.}
\begin{center}
\renewcommand{\arraystretch}{1.2 }
\begin{tabular}{lcccc} \hline\hline
\multicolumn{2}{c}{~~~Model~~~} & $~~~\rho_{xx}$~~~ & ~~~$\rho_{yy}$~~~ & ~~~$\rho_{zz}$~~~ \\ \hline
 ~~~~~~A~~~~~~ & P/AP-Mn$_{\rm Cu}$ & 104  &  71 & 147 \\
 ~~~~~~B~~~~~~ & DLM-Mn$_{\rm Cu}$& ~111~  & ~111~ & ~171~ \\
 ~~~~~~C~~~~~~ & P/AP-Mn$\leftrightarrow$Cu swap& 124  &  97 & 267 \\
 ~~~~~~D~~~~~~ & DLM-Mn$\leftrightarrow$Cu swap& 124  & 124 & 287 \\
 ~~~~~~E~~~~~~ & Cu$_{\rm Mn}$&  24  &  24 & 121 \\
 ~~~~~~F~~~~~~ & Mn-vacancies&  36  &  36 & 155 \\
\hline
 ~~~~~~G~~~~~~ &\tcb{SDR} for Mn$_{\rm Cu}$& 234  & 234 & 363 \\
\hline\hline \end{tabular}
\renewcommand{\arraystretch}{1.2 } \end{center} \label{t4} \end{table}

The following conclusions can be done:
(i) The resistivity in the $z$-direction for all models (the $\rho_{zz}$ component)
is much larger than that in the ($x$,$y$)-plane;
(ii) The in-plane resistivity for the DLM state is more symmetric, i.e.,
$\rho_{xx}$ and $\rho_{yy}$ components are the same while for the collinear
P/AP alignment they are different, because the presence of ordered
moments on a non-magnetic sublattice lowers the symmetry of the system;
(iii) Resistivities roughly follow linear concentration dependence, so one
can say that defect concentrations between 3.5\% to 5\% can reproduce
the experimental (planar) resistivity of 90~$\mu\Omega$cm for Mn$_{\rm Cu}$
or Mn$\leftrightarrow$Cu swap defects while much larger defect concentrations
are needed for Cu$_{\rm Mn}$ defects or Mn-vacancies (no VBS state at the Fermi energy);
(iv) The resistivity for Mn$\leftrightarrow$Cu swaps is slightly larger as
compared to that for Mn$_{\rm Cu}$ defect for the same defect concentrations
due to extra scattering at Cu$_{\rm Mn}$ defects forming the Mn$\leftrightarrow$Cu swap.
The Matthiessen rule is violated, namely, the sum of resistivity for
Models B and E is $\rho_{xx}$= 135~$\mu\Omega$cm while for the Model D it is
124~$\mu\Omega$cm; and
(v) The effect of vertex corrections is small.

We have also tested the effect of electron correlations in LDA+U model.
As an example we have chosen the Model B and the Hubbard parameter $U$=2~eV.
Calculated resistivity components are larger due to the larger scattering on
Mn$_{\rm Cu}$ defects which in turn is due to the increase of the local
Mn-moments caused by correlations.
For example, for Model B we have an increase of $\rho_{xx}$ by about
18~$\mu\Omega$cm, or by 15\%.
We have also considered the phase II.
As an example we have again calculated resistivity for the Model B.
The resistivity is smaller ($\rho_{xx}$=42~$\mu\Omega$cm vs 111~$\mu\Omega$cm
for the phase I case) due to the smaller effective scattering (smaller local
Mn$_{\rm Cu}$-moments).

As an example of the effect of temperature on transport properties we have
calculated \cite{sdr-our} the resistivity in the paramagnetic (DLM) state
above the N\'eel temperature (often called the spin-disorder resistivity, SDR).
The experiment \cite{cumnas-1} indicates a large increase of the planar
resistivity from about 90~$\mu\Omega$cm at 5~K to about 160~$\mu\Omega$cm at 300~K.
Such a large increase cannot be ascribed only to phonons (e.g., a phonon
contribution to the resistivity of about 25~$\mu\Omega$cm exists for bcc-Fe
at the Curie temperature $T=1050$~K).
The largest part of contribution to the resistivity of bcc-Fe should be
ascribed to spin-fluctuations.
\tcb{ To illustrate the effect also for CuMnAs, one can compare the Model B
(Mn-impurity on Cu with the spin-disorder decribed by the DLM state) and
the Model G describing the paramagnetic state (SDR) in which the
spin-disorder exists also on the native Mn-sublattices.
The calculated SDR} is around 230~$\mu\Omega$cm, which looks
reasonably because the temperature of measurement (300~K) is appreciably
smaller than the N\'eel temperature (480~K).
The contribution due to spin-fluctuations monotonically increases with temperature
up to the N\'eel temperature and then remains constant so that the calculated
SDR seems reasonable.

The resistivity depends on the actual occupation of sublattices
which is a challenging problem connected with similar scattering
crossections of atoms forming the alloy.
In Ref. ~\onlinecite{cumnas-str} for a sample grown on GaAs(001) was
suggested that the Cu-lattices are fuly occupied by Cu-atoms while
8\% of Cu- and 8\% of Mn-atoms are found on the As-sublattice leaving
about 14\% vacancies on the Mn-sublattice.
A recent analysis \cite{carlos} for a sample grown on GaP(001) as
in Ref. ~\onlinecite{cumnas-1} has indicated the presence of 10\%
vacancies on both the Cu- and Mn-lattices.
It should be emphasized that actual compositions should not be
taken literally as they depend on the annealing and can also slightly
fluctuate from sample to sample.
Calculated longitudinal resistivities for GaAs and GaP grown samples
are about 180~$\mu \Omega$cm and 88~$\mu \Omega$cm, respectively,
indicating a better agreement with experiment \cite{cumnas-1} for
samples grown on GaP.

\subsection{Exchange interactions and the N\'eel temperature} \label{EI-cumnas}

\begin{figure}[h] \center
\includegraphics[width=6.5cm]{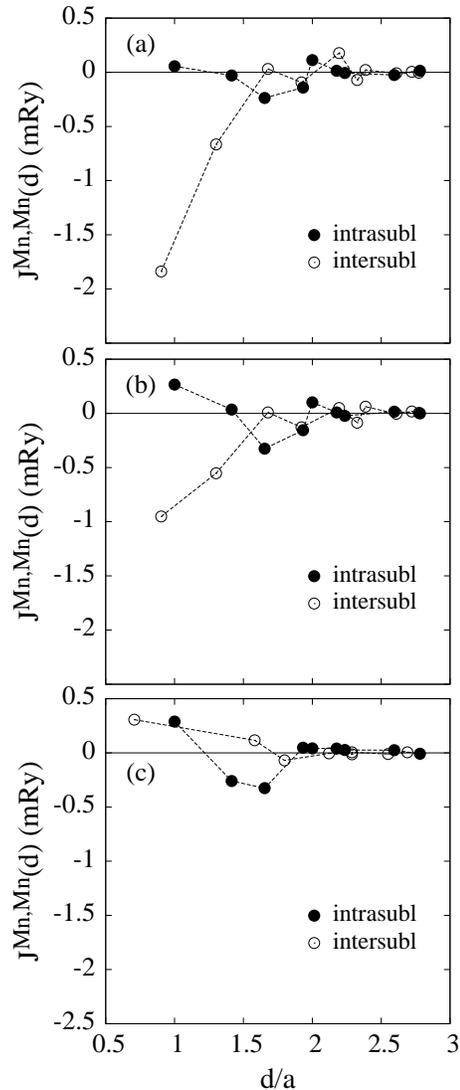} \caption { Exchange interactions for the tetragonal CuMnAs
between Mn-atoms as a function of the distance $d$ (in units of the lattice constant $a$): (a) The
phase I, AFM state, (b) the phase I, paramagnetic (DLM) state, and (c) the phase II, DLM state.
The exchange interactions are subdivided into two groups, namely, between Mn-atoms on the same
sublattice (intrasubl) and between atoms on different sublattices (intersubl). }
\label{f4}
\end{figure}
The exchange interactions in the ideal (defect-free) phase I for both AFM- and DLM-
(paramagnetic) reference states are shown in Figs.~\ref{f4}a and \ref{f4}b,
respectively while the corresponding interactions for the phase II and assuming
the DLM-reference state are shown in Fig.~\ref{f4}c.
In all cases we show interactions among atoms on the same Mn-sublattice
(intrasublattice interactions) as well as among atoms on different Mn-sublattices
(intersublattice interactions).
All other interactions are zero.
The following conclusions can be done:
(i) Exchange interactions for the AFM reference state exhibit, as expected,
a strong leading AFM intersublattice couplings while the intrasublattice ones
are much smaller in their absolute values;
(ii) More important are interactions derived from the paramagnetic (DLM) state.
The paramagnetic state assumes no specific magnetic order and the character
of such interactions is a precursor of the possible AFM ground state (see also
Ref. ~\onlinecite{cumnsb-fe}).
The fact that qualitative character of both intersublattice and intrasublattice
interactions is the same as in the AFM reference state can be interpreted as a
strong indication of the AFM ground state; and
(iii) Because dominating (AFM-like) intersublattice interactions in the
paramagnetic state are smaller than those derived from the AFM reference
state, one can expect a lower N\'eel temperature derived from the DLM state.
We have also estimated exchange interactions for the paramagnetic state
for the phase II which are shown in Fig.~\ref{f4}c.
The interactions are very different, in particular the intersublattice ones.
Consequently, one can expect very different N\'eel temperature.

\begin{figure}
\center \includegraphics[width=8cm]{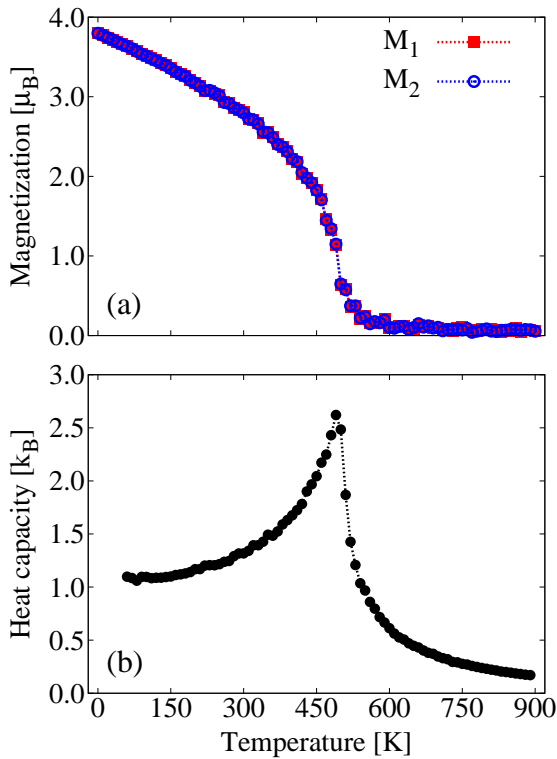}
\caption {(Color online) (a) Magnetizations of the Mn-sublattices as a function
of temperature assuming exchange interactions derived from the paramagnetic (DLM)
state of the tetragonal CuMnAs with the phase I structure.
By symmetry, the dependence of both sublattice magnetizations on the temperature
is the same.
(b) The temperature dependence of the heat capacity from which the N\'eel
temperature can be extracted more accurately (about 480~K). }
\label{f5}
\end{figure}
The N\'eel temperatures were determined using the atomistic spin dynamics
(ASD) codes \cite{asd} which contain the package for the estimate of critical
temperatures using the Monte Carlo simulations.
We show in Fig.~\ref{f5} the sublattice magnetizations and the heat capacity
as a function of the temperature.
The sublattice magnetizations at $T=0$~K (equal to the local Mn-moments) are reduced
with temperature due to spin-fluctuations and disappear at the N\'eel temperature.
An internal test of consistency of calculations is that temperature dependence
of both sublattice magnetizations should be identical and they indeed are.
The magnetization at the N\'eel temperature is not zero, but rather smeared out
due to the finite size of sampling supercells used in the Monte Carlo
calculations.
It should be noted, however, that the zero sublattice magnetizations do not mean
that also local moments are zero.
On the contrary, the local Mn-moments in the AFM- and DLM-states are very similar
due to the rigidity of Mn-moments with respect to their rotations.
The N\'eel temperature can be more precisely extracted from the maximum of the
heat capacity (see Fig.~\ref{f5}b).
The N\'eel temperature estimated in this way and employing exchange
interactions derived from the paramagnetic (DLM) reference state is
about 480~K.\cite{remark}
This represents a good agreement of calculated and experimental N\'eel
temperatures considering the fact that we have assumed an ideal,
defect-free phase I while the real sample\cite{cumnas-tn} contains
unspecified amount of defects.
The estimated N\'eel temperature for the AFM reference state is
higher, being about 680~K, as expected from larger values of exchange
interactions (Fig.~\ref{f4}a vs Fig.~\ref{f4}b).
Finally, we have obtained a paramagnetic state using the exchange interactions
corresponding to the DLM reference state and the phase II (see Fig.~\ref{f4}c).
On the basis of this result, one can exclude the phase II as a ground state.

\section{Conclusions} \label{Con}

We have performed an extensive {\it ab initio} study of electronic, magnetic, and transport
properties of the tetragonal AFM-CuMnAs alloy with potential technological applications.
The VASP approach was used for the phase stability and the estimate of formation energies
of possible defects.
In the next step, for the experimental lattice structure, the TB-LMTO-CPA approach was
adopted to estimate transport properties and the N\'eel temperature from calculated exchange
interactions by the Monte Carlo method.
The main conclusions are:
(i) The theoretical optimized structure of the bulk tetragonal AFM-CuMnAs is the phase II,
but with smaller volume than the experimental one.
The same result was obtained for the experimental lattice parameters and optimized atomic
positions inside the unit cell;
(ii) We have found that electron correlations stabilize the phase I.
(iii) There are indications that the presence of the substrate favors the phase I;
(iv) The presence of various defects even at higher concentrations does not change the
phase preference;
(v) Mn$_{\rm Cu}$, Cu$_{\rm Mn}$, Mn$\leftrightarrow$Cu swaps, and vacancies on
Mn- and Cu-sublattices are defects with low formation energies and thus probable candidates that
can explain the finite sample resistivity;
(vi) Estimated in-plane resistivity of CuMnAs systems with Mn$_{\rm Cu}$ defects and Mn$\leftrightarrow$Cu
swaps for concentrations around 3.5-5\% explains experimentally observed values while
much larger concentrations would be needed for Cu$_{\rm Mn}$ defects or Mn-vacancies.
The origin of larger resistivity can be ascribed to the existence of the well-pronounced
virtual bound state at the Fermi energy for Mn$_{\rm Cu}$ defect or Mn$\leftrightarrow$Cu swap; and
(vii) Estimated N\'eel temperature for ideal, defect-free AFM-CuMnAs agrees reasonably well
with the experiment keeping in mind that sample contains unspecified amount of defects.
On the other hand, the ideal phase II gives a paramagnetic state which
contradicts experimental findings.\\

\begin{acknowledgments}
We acknowledge the financial support from the Czech Science Foundation
(Grant No. 14-37427G) and the National Grid Infrastructure MetaCentrum (project LM2015042) for
access to computation facilities.
\end{acknowledgments}


\begin{thebibliography}{99}
\bibitem{remark} We have tested the robustness of the estimation
of the N\'eel temperature with the respect to the size of the
three-dimensional supercell. The choice $20\times 20\times 20$
gives the N\'eel temperature 500~K.
\end{thebibliography}
\end{document}